\documentclass{webofc}
\usepackage[varg]{txfonts}   
\usepackage{bm}
\usepackage{xcolor}

\newcommand{\wW}{\widehat{W}}

\newcommand{\di}{\mathrm{d}}
\newcommand{\D}{\mathrm{d}}
\newcommand{\wrho}{\widehat{\rho}}
\newcommand{\wT}{\widehat{T}}

\newcommand{\wQ}{\widehat{Q}}

\newcommand{\tr}{\mathrm{tr}}
\newcommand{\Tr}[1]{\mathrm{Tr}\left(#1\right)}
\newcommand{\wP}{\widehat{P}}
\newcommand{\wJ}{\widehat{J}}

\begin{document}
\title{Local equilibrium and Lambda polarization in high energy heavy ion collisions }
%
%

\author{\firstname{Andrea} \lastname{Palermo}\inst{1,2}\fnsep\thanks{\email{andrea.palermo@unifi.it}} \and
        \firstname{Francesco} \lastname{Becattini}\inst{1} 
            \and
        \firstname{Matteo} \lastname{Buzzegoli}\inst{3}
             \and
        \firstname{Gabriele} \lastname{Inghirami}\inst{4}
             \and
        \firstname{Iurii} \lastname{Karpenko}\inst{5}
}

\institute{Università di Firenze and INFN Sezione di Firenze,
Via G. Sansone 1, I-50019 Sesto Fiorentino (Florence), Italy
\and
           Institut f\"ur Theoretische Physik, Johann Wolfgang Goethe-Universit\"at,
Max-von-Laue-Straße 1, D-60438 Frankfurt am Main, Germany
\and
           Department of Physics and Astronomy, Iowa State University,
Ames, Iowa 50011, USA
\and
           GSI Helmholtzzentrum f\"ur Schwerionenforschung GmbH,
Planckstr. 1, 64291 Darmstadt , Germany
\and
           Faculty of Nuclear Sciences and Physical Engineering, Czech Technical University
in Prague, B\v{r}ehov\'a 7, 11519 Prague 1, Czech Republic
          }

\abstract{%
  The polarization of the $\Lambda$ hyperon has become an important probe of the Quark-Gluon Plasma produced in relativistic heavy-ion collisions. Recently, it has been found that polarization receives a substantial contribution from a local equilibrium term proportional to the symmetric derivative of the four-temperature vector, the thermal shear tensor. We show that, at very high energies, this term can restore the agreement between the experimental measurements and the predictions of the hydrodynamic model, provided that the hadronization hypersurface is isothermal. We review the theoretical derivation of this new term, discuss numerical computations at RHIC and LHC energies, and compare them with the experimental data. We also present the effect of feed-down corrections.
}
\maketitle
%
\section{Introduction}
\label{intro}
\let\thefootnote\relax\footnotetext{Presented at the 20th International Conference on Strangeness in Quark Matter (SQM 2022).}
In non-central heavy ion collisions, the Quark-Gluon Plasma (QGP) is produced with a tremendous orbital angular momentum, which in turn polarizes particles produced in the hadronization stage.
The first quantitative relation for the expectation value of the spin-vector of Dirac fermions, was found to be \cite{Becattini:2013fla}:
\begin{equation}\label{eq:spin vort}
     S_\varpi^\mu(p)= -\frac{1}{8m} \epsilon^{\mu\nu\sigma\tau} p_\tau 
 \frac{\int_{\Sigma} \di \Sigma \cdot p \, n_F (1 -n_F) 
 \varpi_{\nu\sigma}}{\int_{\Sigma} \di \Sigma \cdot p \, n_F},
\end{equation}
where $m$ is the mass of the particle, $n_F$ is the Fermi-Dirac distribution function and the integration hypersurface $\Sigma$ is taken to be the decoupling (or freeze-out) hypersurface. Given the four-temperature vector $\beta^\mu=u^\mu/T$ where $u^\mu$ is the four-velocity and $T$ the proper temperature, the antisymmetric thermal vorticity tensor $\varpi_{\mu\nu}$ reads:
\begin{equation}
    \varpi_{\mu\nu}=-\frac{1}{2}(\partial_\mu\beta_\nu-\partial_\nu \beta_\mu).
\end{equation}

The formula \eqref{eq:spin vort} was successful in explaining the data concerning the global polarization, meaning the polarization integrated over all momenta \cite{nature,global} but, unexpectedly, it could not explain the data of polarization as a function of momentum, the so called local polarization \cite{prl local, niida local}. This inconsistency between the global and the local polarization has been known in the literature as the polarization sign puzzle.

Recently, a new contribution coming from the symmetric derivative of the four-temperature was found \cite{shear theory, shear num, yin theory, yin num}. Adopting the results of \cite{shear theory}, the additional term reads: 
\begin{equation}\label{eq:spin shear}
    S_\xi^\mu(p)=-\frac{1}{4m} \epsilon^{\mu\nu\sigma\tau} \frac{p_\tau p^\rho}{\varepsilon}
 \frac{\int_{\Sigma} \di \Sigma \cdot p \; n_F (1 -n_F) 
 \hat{t}_\nu\xi_{\sigma\rho}}{\int_{\Sigma} \di \Sigma \cdot p \; n_F}.
\end{equation}
In the above formula, $\varepsilon=\sqrt{m^2+\bm{p}^2}$ is the particle's energy, the vector $\hat{t}$ is given by $\hat{t}^\mu=(1,0,0,0)$ in the laboratory frame, and its origin will be briefly commented in the next section, and the symmetric tensor $\xi_{\mu\nu}$ is the so-called thermal shear:
\begin{equation}
    \xi_{\mu\nu}=\frac{1}{2}(\partial_\mu\beta_\nu+\partial_\nu \beta_\mu).
\end{equation}

Equation \eqref{eq:spin shear} represents a non-dissipative contribution to spin polarization, because it originates from the local equilibrium statistical operator (see section \ref{sec-1}). Nonetheless, it is a non-equilibrium effect, as in global equilibrium the thermal shear is necessarily vanishing due to the Killing condition. 

To account for both effect, the total spin vector is the sum of equation \eqref{eq:spin vort} and \eqref{eq:spin shear}, that is $S^\mu(p)=S_\varpi^\mu(p)+S_\xi^\mu(p)$. 

\section{Polarization in high energy nuclear collisions: isothermal decoupling}
\label{sec-1}
We comment on some noteworthy steps of the theoretical derivation, having in mind applications to heavy ion collisions at very high energy, e.g $\sqrt{s_{NN}}\geq 200$ GeV. More details about the derivation can be found in \cite{shear theory, shear num}.

The polarization vector is computed using the formula \cite{beca libro}:
\begin{equation}\label{eq:PolVec}
S^\mu(k)=\frac{1}{2}\frac{\int_\Sigma \D\Sigma \cdot k\,\tr\left[\gamma^\mu\gamma^5 W^+(x,k)\right]}
    {\int_\Sigma \D\Sigma\cdot  k\, \tr\left[W^+(x,k)\right]},
\end{equation}
where the integration is performed on the decoupling hypersurface. The function $W^+(x,k)$ is the particle component of the so-called Wigner function, to which we can associate a Wigner operator $\wW$ in such a way that the Wigner function is its expectation value, $W(x,k)=\Tr{\wrho \;\wW(x,k)}$. The particle component of the Wigner operator is:
\begin{equation}\label{eq:wigop}
\widehat{W}^+_{ab}(x,k) =  \theta(k^0)\theta(k^2)\frac{1}{(2\pi)^4} \int \di^4 s \; e^{-i k \cdot s}
   :\bar{\Psi}_b (x+s/2) \Psi_a (x-s/2): \, ,
\end{equation}

For what concerns the density operator $\wrho$, it is assumed to be the so-called local equilibrium density operator:
\begin{equation}\label{eq:LE}
\wrho_{LE}=\frac{1}{Z}e^{-\int\di\Sigma_\mu \wT^{\mu\nu}\beta_\nu},
\end{equation}
where $\wT^{\mu\nu}$ is the stress-energy tensor operator, the integration hypersurface is the decoupling hypersurface and we have neglected the presence of conserved charges. It is known that the density operator \eqref{eq:LE} describes an ideal fluid out of equilibrium, hence the results obtained from \eqref{eq:LE} are non-dissipative \cite{vWeert, zubarev}. 

In the hydrodynamic regime, when dealing with expectation values of local operators such as the Wigner operator $\wW(x,k)$, a Taylor expansion of the four-temperature about the point $x$ is possible, and the density operator is approximated as:
\begin{equation}\label{eq:rho sviluppato}
\wrho_{LE}=\frac{1}{Z_{LE}}\exp\left[-\beta(x)\cdot\wP +\frac{1}{2}\varpi_{\nu\lambda}(x)\wJ_x^{\nu\lambda}-\frac{1}{2}\xi_{\nu\lambda}\wQ_x^{\nu\lambda}\right].
\end{equation}
The thermal vorticity couples with the angular momentum operator $\wJ^{\lambda\nu}$, whereas the termal shear couples with the operator $\wQ^{\lambda\nu}$:
\begin{equation}
\begin{split}
\wJ^{\lambda\nu}_x=&\int_\Sigma \di \Sigma_\mu \; 
  \left[(y-x)^\lambda \wT^{\mu\nu}(y) - (y-x)^\nu \wT^{\mu\lambda}(y)\right],\\
 \wQ^{\lambda\nu}_x =& \int_\Sigma \di \Sigma_\mu \; \left[(y-x)^\lambda \wT^{\mu\nu}(y) + 
  (y-x)^\nu \wT^{\mu\lambda}(y)\right]=\int_\Sigma\di\Sigma_\mu \; \widehat{\mathcal{Q}}^{\mu,\lambda\nu}.\label{eq:Q is not conserved}
\end{split}
\end{equation}
The remarkable difference between $\wJ^{\lambda\nu}_x$ and $\wQ^{\lambda\nu}_x$ is that the latter depends on the integration hypersurface, because $\partial_{\mu}\widehat{\mathcal{Q}}^{\mu,\lambda\nu}\neq 0$. That's why the vector $\hat{t}$ appears in \eqref{eq:spin shear}: $\hat{t}$ is to be interpreted as the average normal vector to the decoupling hypersurface.

If we use linear response theory up to first order in gradients with the operator \eqref{eq:rho sviluppato}, we obtain the results \eqref{eq:spin vort} and \eqref{eq:spin shear}. For very high energy collisions, however, we can do better. In fact, if the collision happens at very high energy the chemical potential is negligible, and the only intensive thermodynamic parameter is the temperature $T$. It ensues that the only possible parametrization of the freeze-out hypersurface is $T=T_\text{dec}$ for some constant decoupling temperature $T_\text{dec}$. Therefore, since $\beta^\mu=u^\mu/T$, \eqref{eq:LE} reduces to the iosthermal local equilibrium density operator:
\begin{equation}\label{eq:rho ILE}
\wrho_{ILE}=\frac{1}{Z}\exp\left[-\int_{T=T_{\text{dec}}} \di\Sigma_\mu\wT^{\mu\nu}\beta_\nu\right]
\simeq 
\frac{1}{Z}\exp\left[-\frac{1}{T_\text{dec}}\int_{T=T_{\text{dec}}} \di\Sigma_\mu\wT^{\mu\nu} u_\nu\right].
\end{equation}

Although $\wrho_{LE}$ and $\wrho_{ILE}$ are completely equivalent in an isothermal decoupling scenario, linear response theory leads to different results depending on which operator we use. On the one hand equation \eqref{eq:rho sviluppato} is expanded in terms of gradients of the four-temperature vector $\beta^\mu$. On the other, for the operator \eqref{eq:rho ILE}, the expansion in gradients would involve only the gradients of $u^\mu$ and not of the temperature $T$. 
If we could compute polarization exactly, it would be irrelevant to use \eqref{eq:LE} or \eqref{eq:rho ILE} as the density operator for an isothermal decoupling hypersurface, but having to resort to linear response theory it is paramount to choose the best density operator before computing the gradient expansion, and in the case of high energy nuclear collisions, the best choice is \eqref{eq:rho ILE}.

Using the isothermal local equilibrium density operator the mean spin vector reads::
\begin{equation}\label{eq:spin ile}
S_{ILE}^\mu(k)= -\frac{1}{8m} \epsilon^{\mu\nu\sigma\tau} k_\tau 
 \frac{\int_{\Sigma} \di \Sigma \cdot k \, n_F (1 -n_F) 
 \left(\omega_{\nu\sigma}+2\hat{t}_\nu\Xi_{\sigma\rho}\frac{p^\rho}{\varepsilon}\right)}{T_{\text{dec}}\int_{\Sigma} \di \Sigma \cdot k \, n_F},
\end{equation}
where we have defined the kinematic vorticity $\omega_{\mu\nu}$ and the kinematic shear $\Xi_{\mu\nu}$:
\begin{align}
\omega_{\mu\nu}=-\frac{1}{2}(\partial_\mu u_\nu -\partial_\nu u_\mu), 
&&\Xi_{\mu\nu}=\frac{1}{2}(\partial_\mu u_\nu +\partial_\nu u_\mu).
\end{align}

In \cite{shear num} it was found that, for a decoupling temperature $T_\text{dec}\sim155$ MeV, the predictions of \eqref{eq:spin ile} are in very good agreement with the data, thus solving the polarization sign puzzle. Such result confirm that, for high energy applications, the use of \eqref{eq:rho ILE} and \eqref{eq:spin ile} is more appropriate. 

\section{Feed-down corrections}
Hitherto, only the contribution of primary $\Lambda$s has been considered, whereas feed-down may have some significance. We report for the first time the effect of feed-down on polarization including the thermal shear term at $\sqrt{s_{NN}}=200$ GeV and $\sqrt{s_{NN}}=5020$ GeV, corresponding to RHIC and ALICE top energies. 
The full detail of the simulation will appear in a forthcoming publication \cite{prep}. There have been already precious studies on the feed-down correction to polarization, and we refer the reader to them for a detailed theoretical analysis \cite{xia, beca fd}.

Considering only the main channels of production of $\Lambda$ hyperons, namely the primary $\Lambda$s, and those produced in the processes $\Sigma^*\rightarrow\Lambda +\pi$ and $\Sigma_0\rightarrow \Lambda + \gamma$, total spin vector of the $\Lambda$ particle in its rest frame  is given by:
\begin{align}\label{eq:tot spin decay}
    \bm{S}_{\Lambda,tot}(p)=\frac{n^{(FO)} \bm{S}^{(FO)}(p)+n^{(\Sigma^*)} \bm{S}^{(\Sigma^*)}(p)+n^{(\Sigma_0)}\bm{S}^{(\Sigma_0)}(p)}{n^{(FO)}+n^{(\Sigma^*)}+n^{(\Sigma_0)}}.
\end{align}
where $\bm{S}^{(X)}$ are the spin vector computed in each production channel, where the primaries have been denoted by $(FO)$, and the other channels have the label of the mother particle. The vector $\bm{S}^{(FO)}$ corresponds to the back-boost of \eqref{eq:spin ile} to the rest frame of the $\Lambda$, and $\bm{S}^{(\Sigma^*)}$ and $\bm{S}^{(\Sigma_0)}$ are calculated using the formula (31) of ref. \cite{beca fd}.

In equation \eqref{eq:tot spin decay}, $\ n^{(X)}$ represent the fraction of the total $\Lambda$s produced in the channel $X$. An estimation has been given in \cite{beca fd}, namely $n^{(FO)}_\Lambda=0.243$, $n^{(\Sigma^*)}_\Lambda=0.359$ and $n_{\Lambda}^{(\Sigma_0)}=0.165$. These fractions are used for both $200$ and $5020$ GeV calculations. 

The hydrodynamic stage of the evolution of the $QGP$ is simulated using the 3+1 D hydrodynamical code vHLLE \cite{Karpenko.2013}. The initial state has been generated with averaged entropy density profile from the Monte Carlo Glauber 
model, generated by GLISSANDO v.2.702 code \cite{Rybczynski.2013}. 

\begin{figure}
\centering
\includegraphics[width=0.45\textwidth]{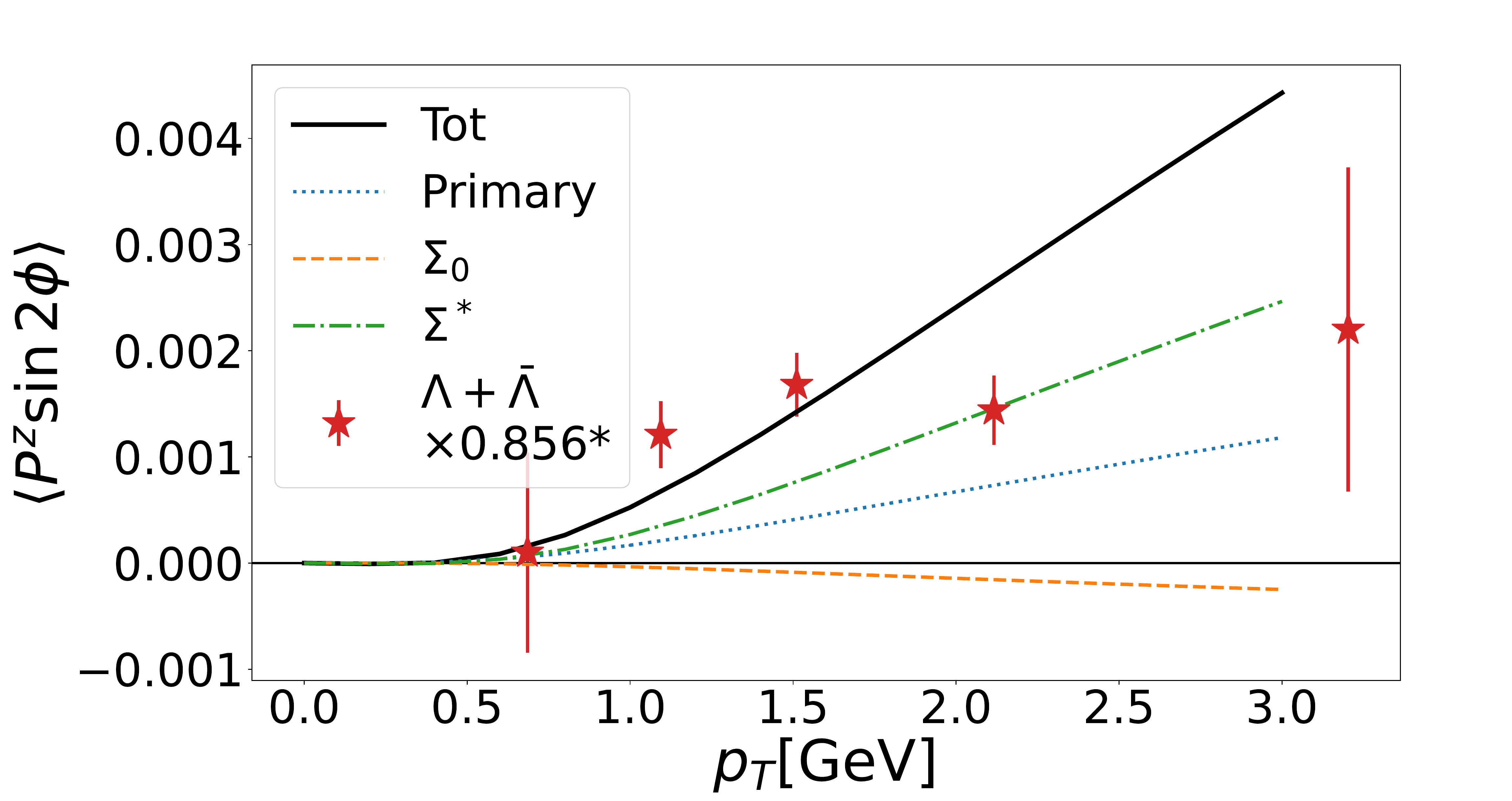}
\includegraphics[width=0.45\textwidth]{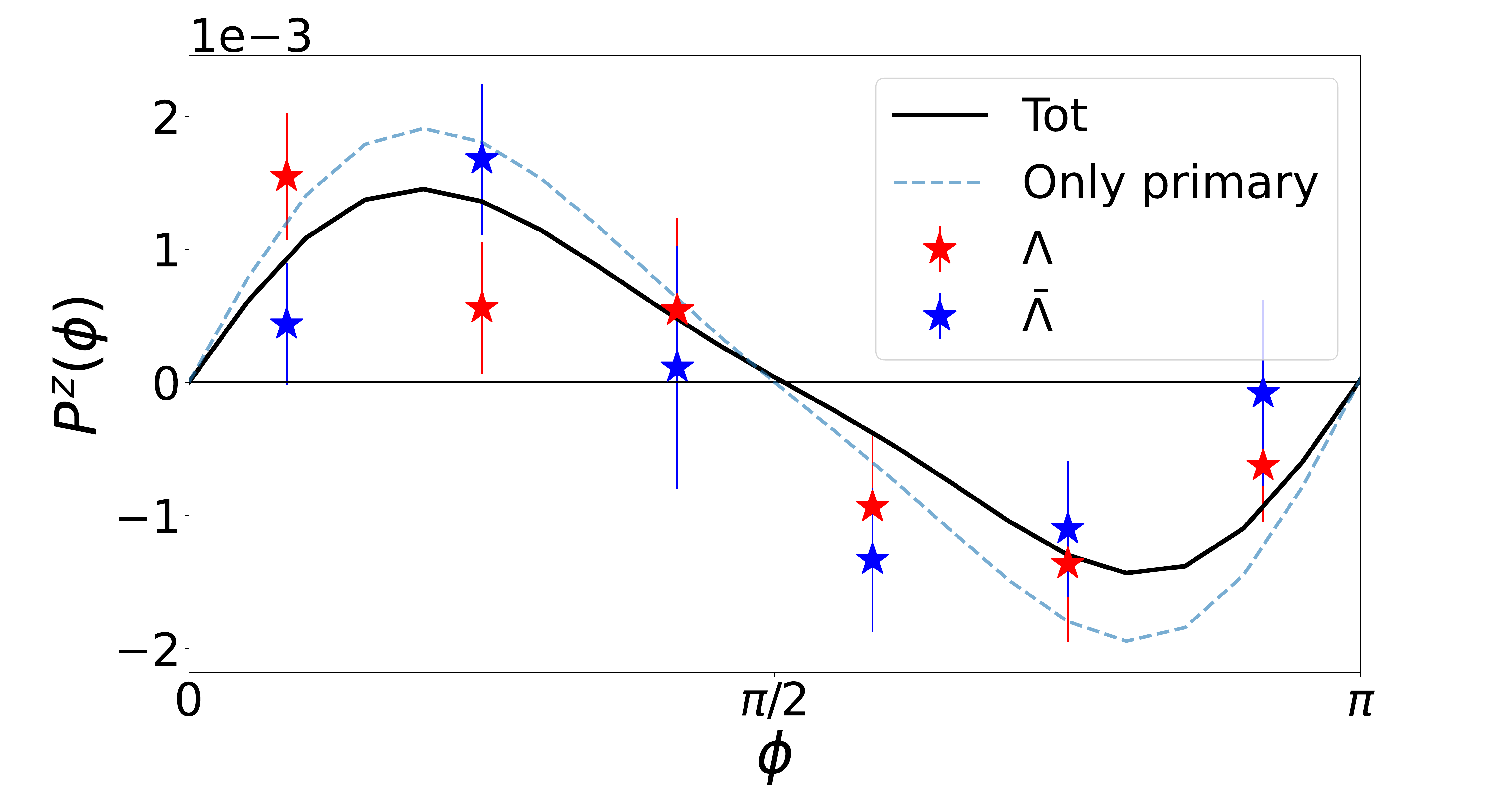}
\caption{Calculations of the polarization vector at $200$ GeV. Left panel: second Fourier harmonic of $P_z$. The contribution of each channel is plotted after multiplication for the respective weight $n^{(x)}$. Right panel: azimuthal dependence of the feed-down corrected $P_z$ compared to the primary-only channel. The data points are taken from \cite{prl local}, and we used $\langle\cos^2\theta^*_p\rangle=1/3$ and $\alpha_H=0.732$. }
\label{fig-1}       
\end{figure}

\begin{figure}
\centering
\includegraphics[width=0.45\textwidth]{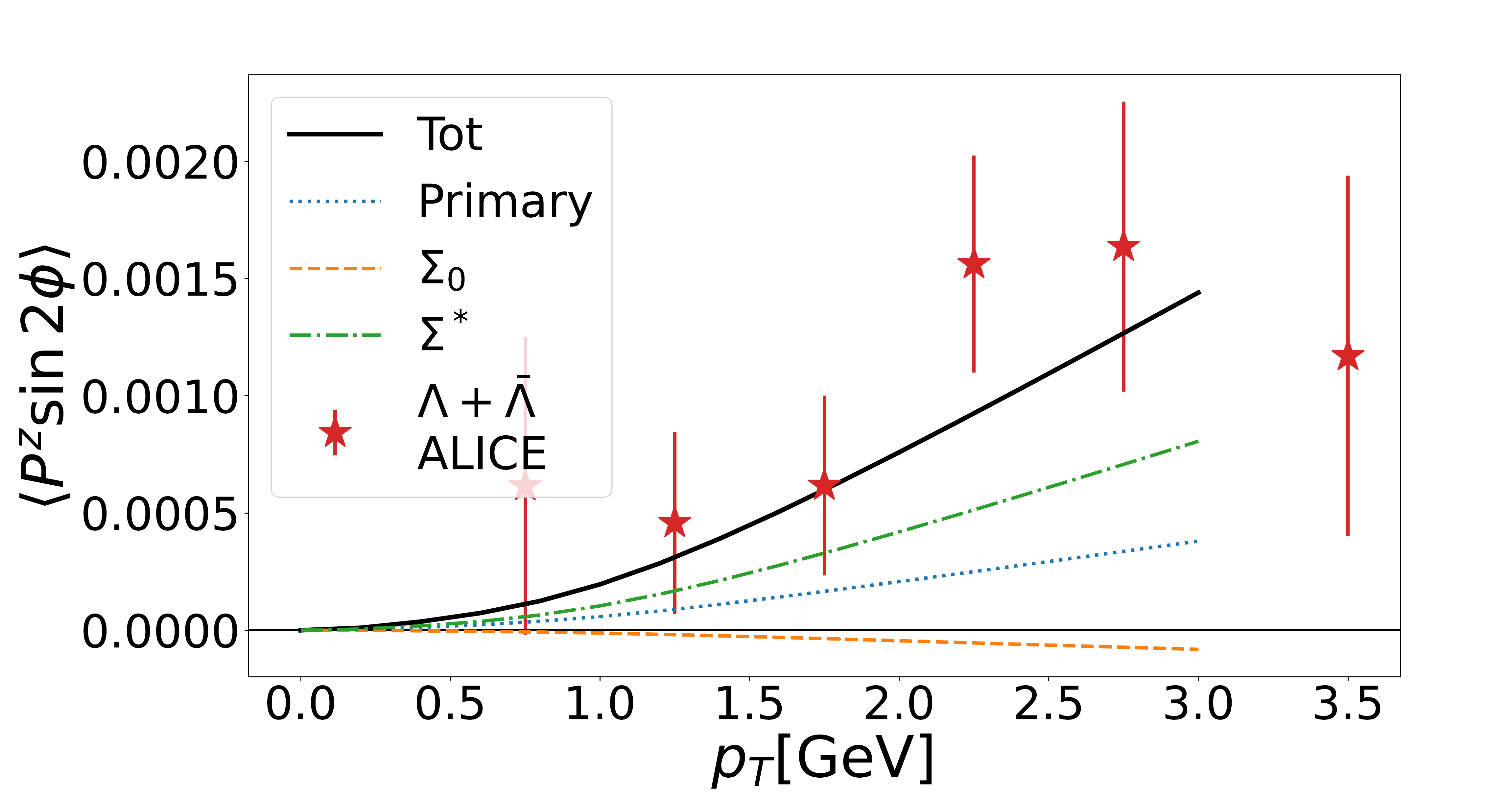}
\includegraphics[width=0.45\textwidth]{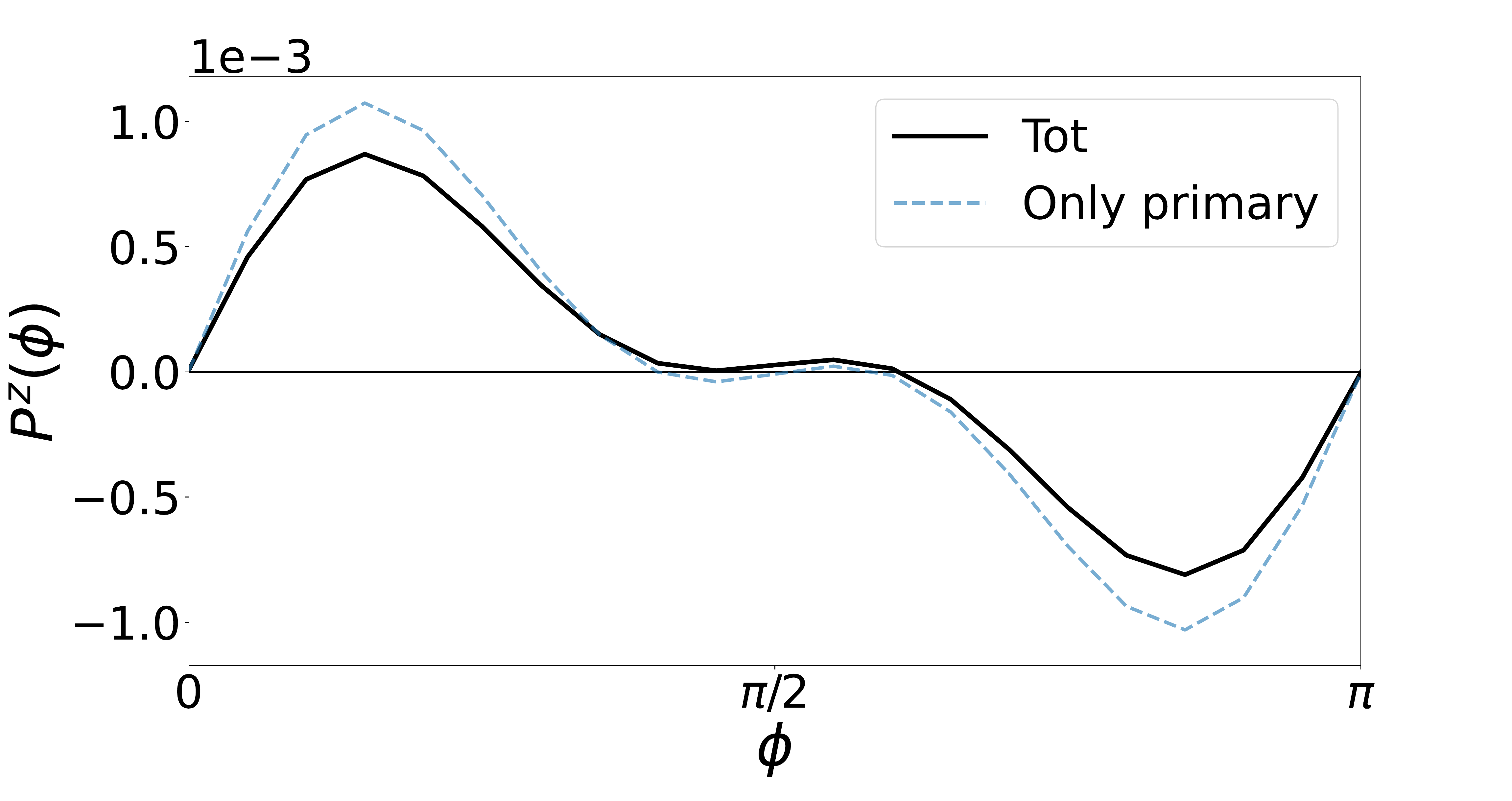}
\caption{Calculations of the polarization vector at $5020$ GeV. Left panel: second Fourier harmonic of $P_z$, similar as figure \ref{fig-1}. The data points are taken from \cite{alice}. Right panel: comparison between $P_z$ before and after the inclusion of feed down, much like fig. \ref{fig-1}. In this simulation a non vanishing bulk viscosity has been used.}
\label{fig-2}       
\end{figure}

In figures \ref{fig-1} and \ref{fig-2}, we show the dependence of $\langle P_z\sin2\phi\rangle$ as a function of $p_T$ and $P_z$ ($z$ being the beam axis) as a function of $\phi$, for $\sqrt{s_{NN}}=200$ GeV and $\sqrt{s_{NN}}=5020$ GeV respectively, using the isothermal decoupling result for the spin vector \eqref{eq:spin ile}. We remind to the reader that the polarization vector for fermions is defined as $P^\mu=2S^\mu$ where $S^\mu$ is the spin vector.

A good agreement with the data is found for a decoupling temperature of $T_\text{dec}=160$ MeV. For $200$  GeV calculations, the simulation refers to $20-60$\% centrality, and we used $\eta/s=0.08$ and $\zeta/s=0$ where $\eta$ and $\zeta$ are the shear and bulk viscosities and $s$ is the entropy density. The simulation at $5020$ GeV, on the other hand, refers to $30-50$\% centrality and $\eta/s=0.12$. Interestingly, the experimental data cannot be reproduced if we keep $\zeta/s=0$, so that a non-vanishing bulk viscosity is used. In fig. \ref{fig-2}, the results are reported using  $\zeta/s$ as in the parametrization III of ref. \cite{bulk}.

\label{sec-3}

\section{Conclusion and outlook}
\label{this is the end beautiful friend}
We have studied polarization of $\Lambda$ hyperons at local equilibrium, discussing both the contributions coming from thermal vorticity and from thermal shear. We have argued that the isothermal local equilibrium density operator should be used when dealing with high energy nuclear collisions and we have used it to calculate polarization including the feed-down correction. The feed-down reduces the signal of about $20\%$ compared to the case of primary $\Lambda$s and a good fit with the data is obtained for a decoupling temperature of $160$ MeV. We have found that, at $5020$ GeV a non vanishing bulk viscosity is needed to explain the experimental data. The origin of this effect and the importance of bulk viscosity for spin physics represent an interesting avenue for future research.

\end{document}